\documentclass[manuscript]{aastex631}

\received{}
\revised{}
\accepted{}

\shorttitle{S-stars}
\shortauthors{A. Burkert et al.}

\begin{document}

\title{The Orbital Structure and Selection Effects of the Galactic Center S-Star Cluster}

\author{A. Burkert}
\affiliation{University Observatory Munich (USM) \\ Scheinerstrasse 1, 81679 Munich, Germany}
\affiliation{Max-Planck-Institut f\"ur extraterrestrische Physik (MPE) \\ Giessenbachstr. 1, 85748 Garching, Germany}

\author{S. Gillessen}
\affiliation{Max-Planck-Institut f\"ur extraterrestrische Physik (MPE) \\ Giessenbachstr. 1, 85748 Garching, Germany}

\author{D. N. C. Lin}
\affiliation{Department of Astronomy and Astrophysics, University of California \\ Santa Cruz, USA}
\affiliation{Institute for Advanced Studies, Tsinghua University \\ Beijing, China}

\author{X. Zheng}
\affiliation{Beijing Planetarium, Beijing Academy of Science and Technology \\ Beijing, China}

\author{P. Schoeller}
\affiliation{ORIGINS Excellence Cluster \\ Boltzmannstr. 2, 85748 Garching, Germany}

\author{F. Eisenhauer}
\affiliation{Max-Planck-Institut f\"ur extraterrestrische Physik (MPE) \\ Giessenbachstr. 1, 85748 Garching, Germany}

\author{R. Genzel}
\affiliation{Max-Planck-Institut f\"ur extraterrestrische Physik (MPE) \\ Giessenbachstr. 1, 85748 Garching, Germany}

\correspondingauthor{A. Burkert}
\email{burkert@usm.lmu.de}

\newcommand\msun{\rm M_{\odot}}
\newcommand\lsun{\rm L_{\odot}}
\newcommand\msunyr{\rm M_{\odot}\,yr^{-1}}
\newcommand\be{\begin{equation}}
\newcommand\en{\end{equation}}
\newcommand\cm{\rm cm}
\newcommand\kms{\rm{\, km \, s^{-1}}}
\newcommand\K{\rm K}
\newcommand\etal{{\rm et al}.\ }
\newcommand\sd{\partial}

\begin{abstract}
The orbital distribution of the S-star cluster surrounding the supermassive black hole in the center of the Milky Way is analyzed. A tight, roughly exponential dependence of the pericenter distance r$_{p}$ on orbital eccentricity e$_{\star}$ is found, $\log ($r$_p)\sim$(1-e$_{\star}$), which cannot be explained simply by a random distribution of semi-major axis and eccentricities. No stars are found in the region with high e$_{\star}$ and large log r$_{p}$ or in the region with low e$_{\star}$ and small log r$_{p}$. G-clouds follow the same correlation. The likelihood P(log r$_p$,(1-e$_{\star}$)) to determine the orbital parameters of S-stars is determined. P is very small for stars with large e$_{\star}$ and large log r$_{p}$. S-stars might exist in this region. To determine their orbital parameters, one however needs observations over a longer time period. On the other hand, if stars would exist in the region of low log r$_{p}$ and small e$_{\star}$, their orbital parameters should by now have been determined. That this region is unpopulated therefore indicates that no S-stars exist with these orbital characteristics, providing constraints for their formation. We call this region, defined by $\log$ (r$_p$/AU) $<$ 1.57+2.6(1-e$_{\star})$, the zone of avoidance. Finally, it is shown that the observed frequency of eccentricities and pericenter distances is consistent with a random sampling of log r$_{p}$ and e$_{\star}$. However, only if one takes into account that no stars exist in the zone of avoidance and that orbital parameters cannot yet be determined for stars with large r$_{p}$ and large e$_{\star}$.
\end{abstract}


\section{Introduction}
The galactic center hosts one of the most mysterious structures of the Milky Way, the S-star cluster (SSC; see \cite{genzel2010} for a review). This spheroidal ensemble of at least 50 young, preferentially B-type stars with masses in the range of 8-20 M$_{\odot}$ and 
ages of less than 15 Myrs \citep{gillessen2017, habibi2017}
is confined to a region of 0.04 pc from Sgr A$^*$\citep{genzel2003, ghez2003b, gillessen2009}.  
Their eccentric to nearly parabolic orbits have been reconstructed based on proper 
motion measurements \citep{Genzel1997, genzel2010, Ghez1998, Ghez2003a, ghez2008, 
gillessen2009, schodel2009, boehle2016}.
This kinematic data not only  confirms the existence of a super-massive 
black hole (SMBH) with a mass $M_{\rm SMBH} = 4.297  \times 10^6 M_\odot$ 
at the Galactic Center \citep{Genzel1997, Ghez1998, schodel2002, ghez2005, 
eisenhauer2005}. It also provides valuable clues on the dynamical 
evolution of these recently formed stars in its proximity \citep{levin2003}.

The origin of the SSC is puzzling as single stars typically condense out of molecular 
cloud cores with radii of order 0.1 pc \citep{hester1996, mellema2006, forbes2021}. This is 
larger than the whole SSC. Therefore, the S-stars formed under very exotic conditions
\citep{Goodman2003, thompson2005, Nayakshin2007, levin2007} or were transported 
into this tiny, hostile environment \citep{syer1991, artymowicz1993, davies2020} of 
the central SMBH. If the S-stars formed in situ, 
some violent star formation event must have happened very close to the SMBH a few 
of Myrs ago \citep{morris1993, ghez2003b}. Currently, the S-stars are embedded in a 10$^6$K hot, 
diffuse gas bubble \citep{xu2006, murchinkova2019} which is not a favorite condition for star formation. One should, however, note 
that not all of the gas is hot. A cluster of cold G-clouds with temperatures of order 10$^4$K
has been discovered in the same region as the SSC \citep{gillessen2012, gillessen2013, 
pfuhl2015, plewa2017}. These gas clumps have very small masses, of order several Earth masses. Their nature and origin are still a matter of debate \citep{ burkert2012, murrayclay2012, 
scoville2013, ballone2013, owen2023}
and at the moment, it is not clear whether they are related to the origin of the S-stars.

Today, nearly 200 massive stars have been detected in the Sgr A$^*$  proximity.  In addition
to the SSC, $\sim 20 \% $ of the known stars belong to a subgroup of O/WR stars with top-heavy luminosity function and estimated age
$\sim 6 \pm 2$ Myr \citep{ghez2003b}. They reside in a possibly-warped disk with clock-wise rotation, a semi-major axis of $0.04 {\rm pc} \lesssim a_\star \lesssim 0.5$ pc and an eccentricity of $e_\star \sim 0.2-0.4$, commonly referred to as the clockwise disk (CWSs, \cite{levin2003, paumard2006, lu2006, lu2009, lockmann2009, bartko2009, bartko2010, yelda2014}). The disk stars have an inner semi-major-axis
hole of 0.04 pc, corresponding to the sharp outer edge in the semi-major axis 
distribution of the SSC.  This coexistence and the very similar ages provide
evidence that the SSC and its surrounding disk population might have a common origin. The remaining fraction of the young stars are the ``off-disk'' stars 
(ODSs) with much less well-determined, probably-large eccentricities and inclinations, as well as a non-isotropic distribution, surrounding the disk-star population.
Several over-dense concentrations in their orbital angular-momentum orientation distribution have been 
identified, including the controversial identification of a second disk, inclined by $\sim 110^{\circ}$ with respect to the clockwise disk \citep{paumard2006, 
lu2009, yelda2014, ali2020}.  The luminosity function of the ``off-disk'' stars is 
similar to that of the S stars and less top-heavy than the disk stars \citep{vonfellenberg2022}.  
There are suggestions that two disks of massive stars formed 6 Myrs ago, orbiting
the central SMBH from one or more infalling and colliding clouds \citep{bonnell2008, hobbs2009, 
alig2013}.  It is also possible that secular perturbation by an intermediate-mass companion
\citep{zheng+2020, zheng+2021, straub2023} and dynamical relaxation \citep{rauch1996, Freitag2006, Alexander2007, 
Madigan2011, kocsis2011, kocsis2015} heated a common-natal disk, generating all the diverse 
kinematic properties in each subgroup, including the spheroidal SSC. We will discuss this scenario in greater detail in a companion paper (Zheng et al. in preparation). 

Another, maybe complementary scenario for the origin of the S-stars is Hill's mechanism, whereby massive stellar binaries, generated in the larger environment of the SMBH were 
deflected by some process onto an orbit that brought them so close to the SMBH that they 
broke up \citep{hills1975, hills1988, gould2003, Yu2003, ginsburg2006, amaroseoane2012}.
Typically the more massive star will lose orbital energy and go into a tight, bound 
orbit around the SMBH while the other star gains energy and might even be ejected as a 
high-velocity star or a hyper-velocity star that is unbound to the Milky Way.
Stars with such high velocities have been detected in the galactic halo. Observations indicate 
that these stars have their origin in the galactic center, consistent with this theory. 
One caveat of the Hills mechanism is, however, the fact that it alone can only explain 
the current orbits of the most bound S-stars. One would still need secular evolution 
scatter processes to account for the S-stars at larger distances. In addition, this mechanism does not provide an answer to the question of why almost all S-stars are B-stars, whereas Hill's mechanism should also work equally well for O-stars.

The size and geometry of the SSC provide insight into its origin. More detailed information is, however, available from measurements of the orbital distribution of its stars. By now, we have up to 25 yrs of uninterrupted monitoring of stellar orbits, and the number of stars with known orbits \citep{schodel2002, ghez2003b} has increased to $\sim$ 50 \citep{eisenhauer2005, ghez2005, gillessen2009, gillessen2017}.
One of the most prominent members of the SSC is S2 which passed its first observed pericenter in 2002 
and again in 2018, providing the most precise measurement of the black hole mass and 
allowing for the first time to detect gravitational redshift \citep{gravity2018, do2019} and Schwarzschild precession \citep{gravity2020}, in excellent agreement with the predictions by general relativity. In order to understand the formation 
of the SSC, one however needs to investigate the physical property of the whole cluster, which requires a detailed understanding of biasing and selection effects.

In this paper, we focus on the distribution of the orbital parameters of the SSC. 
In section 2, we demonstrate that the S-stars follow a tight correlation between their orbital eccentricity e$_{\star}$ and their pericenter distance r$_{p}$. Interestingly the G-clouds follow the same distribution, which might indicate a common origin. In section 3, we show that this correlation differs strongly from a thermal distribution which is usually adopted to characterize the SSC or other theoretical models that explain the SSC as the result of secular heating of a previous disk of stars. We also demonstrate that the lack of stars with high e$_{\star}$ and large r$_{p}$ can be understood as a result of selection effects. Stars could indeed exist in this regime. One would, however, need observations on a longer timeline to determine their orbital parameters. The zone of avoidance, on the other hand, is real. It is defined by a sharp edge of r$_{p}$ versus e$_{\star}$, below which no S-stars exist. A summary follows in section 4.

\section{The correlation between orbital eccentricity and pericenter distance}

\begin{figure}
\begin{center}
\includegraphics[width=14.cm]{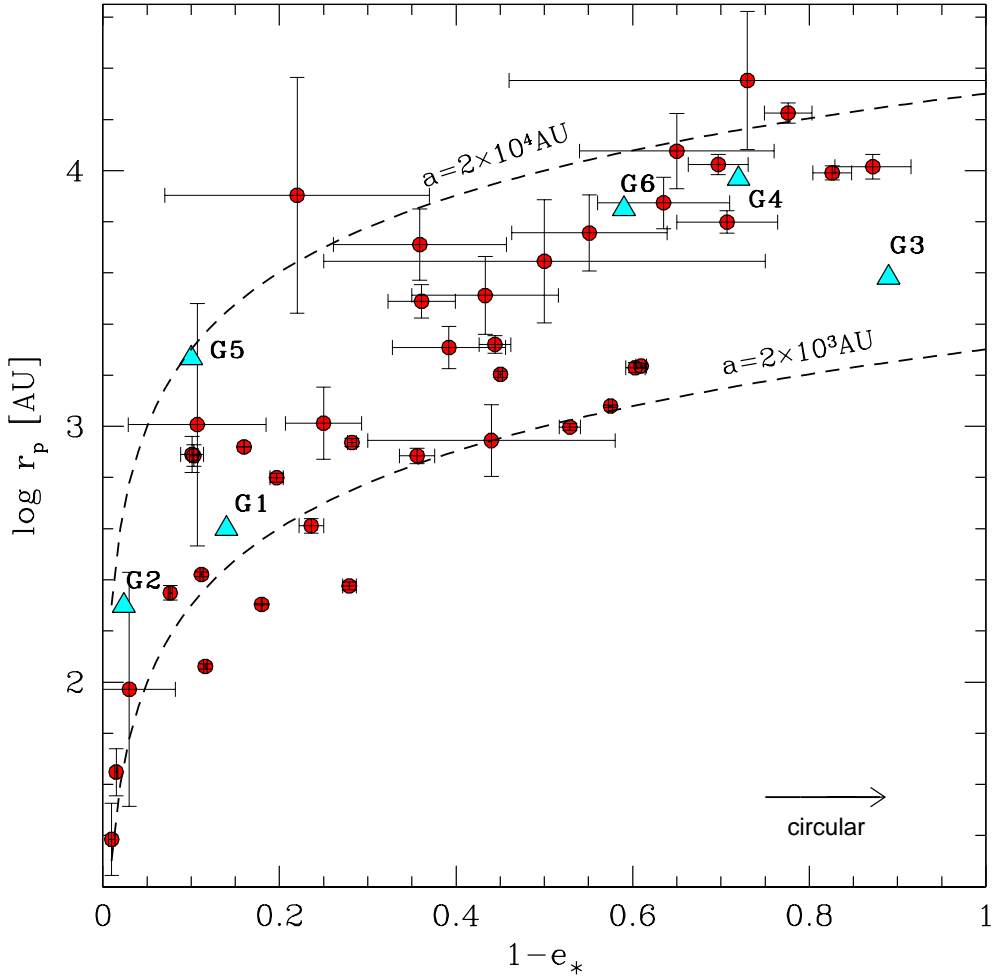}
\caption{Pericenter distances r$_p$ of S-stars (red points) and G-clouds (cyan triangles) versus their orbital eccentricities e. More circular orbits, corresponding to larger values of (1-e), have larger pericenter distances. Dashed lines show the expected dependence of r$_p$ on (1-e) for semi-major axi of $2\times 10^3$AU and $2\times 10^4$AU, respectively. Stars and clouds with larger r$_p$, on average, have larger values of a than objects with smaller r$_p$. A random distribution of semi-major axi, independent of e, therefore cannot explain the data.}
\label{fig:fig1}
\end{center}
\end{figure}

The red points with error bars in figure 1 show the logarithm of r$_{p}$ versus e$_{\star}$ for the S-stars, compiled by \cite{gillessen2017}. The pericenter increases roughly exponentially with increasing (1-e$_{\star}$). In addition to the stars, cyan triangles mark the currently known G-gas clouds \citep{gillessen2012, eckart2013, plewa2017, ciulo2020, peissker2020}
which follow the same correlation closely. This might indicate a common origin of S-stars and G-clouds or a common secular redistribution process that generated the currently observed configuration. Note however that scattering processes should be more disruptive for the G-clouds if they are diffuse, pressure-confined gas clumps, with negligible self-gravity \citep{ burkert2012}. 

One might expect a dependence of r$_p$ on e$_{\star}$ if all the stars have a similar semi-major axis, and a spread in e$_{\star}$, as for given a$_{\star}$ the pericenter distance r$_p$ = a$_{\star}$ $\times$ (1-e$_{\star}$) decreases with increasing e$_{\star}$. In this case, r$_{p}$ should however increase linearly with (1-e$_{\star}$), which is different from the approximately exponential correlation, seen in figure 1. The two dashed black lines depict the expected correlation for a semi-major axis of a$_{\star}$=$2\times 10^3$ AU and a$_{\star}$=$2\times 10^4$ AU, respectively. One can see that the more circular the orbits, the larger the semi-major axis.
This is demonstrated more clearly in figure 2, which shows the eccentricity dependence of a$_{\star}$ (left panel) and of the apocenter distance r$_a$=a$_{\star}$ $\times$ (1+e$_{\star}$). Both quantities are not randomly distributed but depend on (1-e$_{\star}$). We, therefore, can conclude that the dependence of pericenter distance on eccentricity is not just a result of a universal semi-major axis, independent of eccentricity.

\begin{figure}
\begin{center}
\includegraphics[width=16.cm]{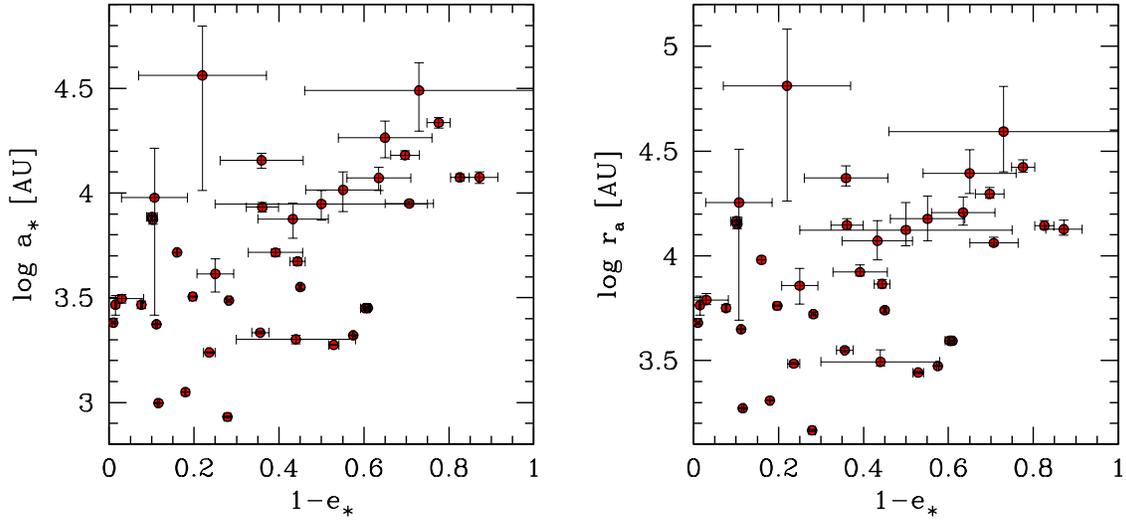}
\caption{Semi-major axis a$_{\star}$ (left panel) and apocenter distance r$_{a}$ of S-stars versus their orbital eccentricities e$_{\star}$. More circular orbits have larger semi-major axis and larger apocenter distances.}
\label{fig:fig2}
\end{center}
\end{figure}

\section {Selection effects}

\noindent One possible explanation of the correlations seen in figure 1 and figure 2 is selection effects. It might, for example,
be harder to determine the complete orbital parameters of stars in the upper left and lower right corners of figure 1.
In order to test this conjecture, we have performed a detailed Monte Carlo study. 
We start by focusing on a pair of r$_{p}$ and (1-e$_{\star}$). Stars are then generated with this set of orbital parameters
on randomly oriented orbital planes. Each star is placed somewhere on its orbit with the starting point at time t$_0$ chosen
randomly between zero and the orbital period. The star is then advanced by 20 years which corresponds to the average observational period and its projected
position (x$_i$,y$_i$) at time t$_i$=t$_0$+(i-1)$\Delta$t in the x-y plane is recorded with $\Delta$t = 1 yr.
Here we assume that the z-axis represents the line of sight to the observer.
Next, a Gaussian random error of 10AU is added to each position which is the expected mean observational error. From this, we determine the time dependence of the distance to the central black hole
r$_i$(t$_i$)=$\sqrt{x_i^2+y_i^2}$.
A second-order regression line r(t)=a+b$\times$t+c$\times$t$^2$ is now fitted through the 21 data points r$_i$(t$_i$).
Repeating this procedure a large number of times allows us to determine the mean value $\langle$c$\rangle$ and its error $\delta c$.
It is this curvature $\langle$c$\rangle$ in the orbit that is required in order to determine the full orbital parameters of a star.
We now follow the selection criterion of \cite{gillessen2017} and assume that the orbital parameters are known if 
$\delta c$/$\langle$c$\rangle \leq 10^{-3}$. The fraction of cases that fulfill this requirement then defines the probability P to determine the orbital parameters for a given of r$_{p}$ and (1-e$_{\star}$). For P=1, orbital parameters can be determined for all configurations.
The smaller P, the less frequently can these parameters be specified. Regions with small values of P should not be populated in figure 1,
even if stars exist on these orbits, because the curvature c and by this, the orbital parameters r$_p$ and e$_{\star}$ cannot be determined reliably for a large fraction of orbital configurations.

We find that, to good approximation, the probability P to measure the orbital parameters of a star with pericenter distance r$_p$ in units of AU and orbital eccentricity e$_{\star}$ is given by the following fit formula

\begin{equation}
log P = -2+ \left(\frac{-q_2 + \sqrt{q_2^2 - 4 q_1 q_3}}{2 q_1} \right)
\end{equation}

\noindent with

\begin{eqnarray}
q_1 = 0.095 \times x \\ \nonumber
q_2 = 1.266 - 0.95 \times x - 0.075 \times x \times \ln x \\
q_3 = \ln (r_p/AU) - 11.4 + 0.855 \times x + 0.075 \times x \times \ln x - \ln x \nonumber
\end{eqnarray}

\noindent and x=(1-e$_{\star}$).

\begin{figure}
\begin{center}
\includegraphics[width=14.cm]{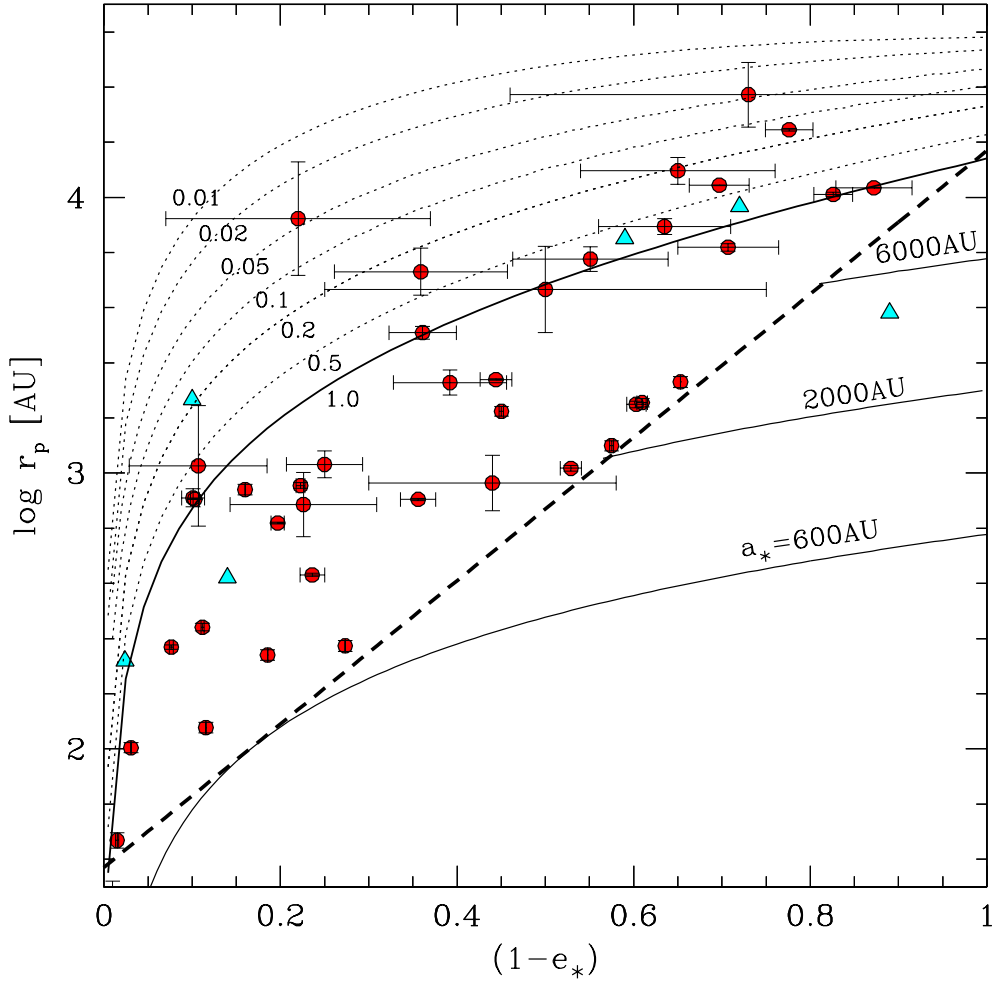}
\caption{Dotted lines show contours of  constant probability P (see labels) to determine the orbits of S-stars within 20 years of observations. It should be possible to
determine the orbital parameters for all stars with pericenter distances below
the solid black line, labeled 1.0. Red points with error bars show those S-stars for which orbital parameters have been determined. Cyan triangles show G-clouds. The dashed black line marks the upper boundary of the zone of avoidance within which no stars with known orbital parameters are found, despite the fact that it should be easy to determine their orbital parameters. For reference, the solid lines in the zone of avoidance show positions of constant semi-major axi a$_{\star}$.}
\label{fig:fig3}
\end{center}
\end{figure}

The dotted lines in figure 3 show contours of constant P. Above a given contour, the probability is smaller than the corresponding value. Red points show the
S-stars with known orbital parameters.  One can see that the unpopulated upper left region with high eccentricities and high r$_p$ is characterized by very small probabilities of a few percent or lower. Indeed, almost no stars lie in this region. And those few stars that are found there have large error bars. We, therefore, cannot rule out that this region is populated with S-stars. However, determining their orbital parameters will require a longer epoch of observations and/or higher observational precision. On the other hand, the orbital parameters of S-stars in the empty lower right triangular region below
the dashed line should by now have been measured. We call this region the zone of avoidance. It is characterized by

\begin{equation}
\log (r_p/AU) \leq 1.57+2.6(1-e_{\star})
\end{equation}

\noindent with the dashed line showing its upper boundary. That this zone of avoidance is not populated by stars therefore indicates that no S-stars exist with these combinations of orbital parameters. The zone of avoidance overlaps with the P=1 line (solid black line) close to e$_{\star}$=0. The orbits of stars on circular orbits above the zone are, therefore, harder to determine as the probability to measure their parameters within on average 20 years is small. This might account for the fact that not many stars with eccentricities close to unity have been reliably detected yet.  

\begin{figure}
\begin{center}
\includegraphics[width=14.cm]{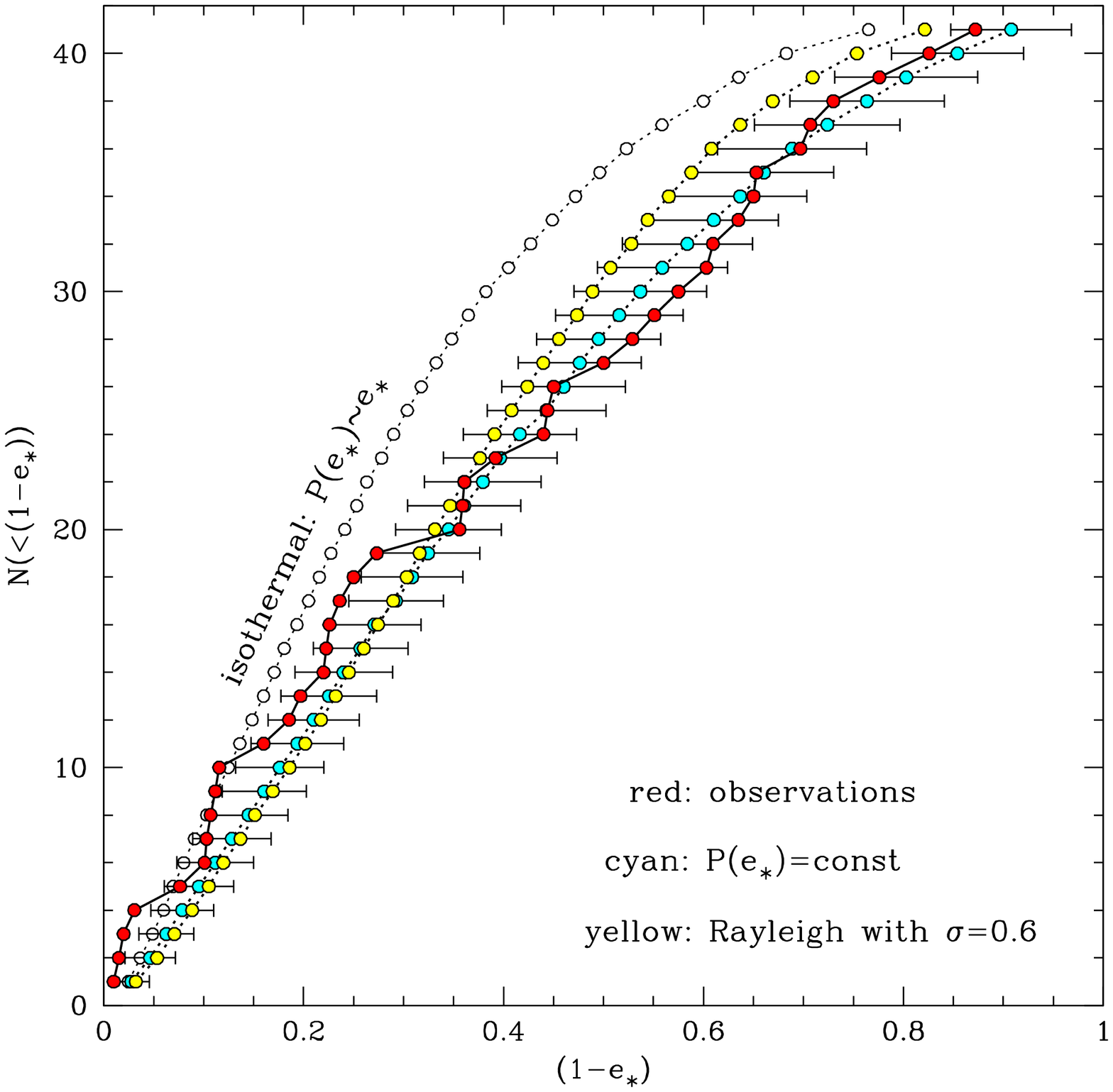}
\caption{Number of S-stars with known orbital parameters and with (1-e$_{\star}$) less or equal to a given value. The red points show the observations. The cyan points with
error bars correspond to a homogeneous distribution in log r$_p$ and e$_{\star}$. Yellow points depict a Rayleigh distribution with $\sigma$=0.6 (see equation 4).  An isothermal distribution with P(e$_{\star}$)$\sim$e$_{\star}$ is shown by the open points. In all cases, stars were only taken if they lie above the zone of avoidance.
In addition, the stars were selected considering the probability of determining their orbits.}
\label{fig:fig4}
\end{center}
\end{figure}

To more quantitatively evaluate the effect of the detection probability on the distribution of observed orbital parameters, figure 4 shows a Monte-Carlo test where we randomly draw values of e$_{\star}$ in the regime of 0 and 1 and values of log(r$_p/$AU) within a range of 1 and 5. We then discard this set of parameters if it lies in the zone of avoidance (equation 3). Otherwise, we accept the set with a probability as given by equation 2. Note that this procedure makes the result insensitive to the exact values of the upper and lower boundary of the adopted log(r$_p$) as long as the lower boundary
lies within the zone of avoidance and the upper boundary lies in the region where the probability of determining the orbital parameters is small. As soon as 41 stars have been selected this way, we determine their cumulative eccentricity distribution. We repeat this procedure 1000 times. The cyan points in figure 4 show the mean of the cumulative e$_{\star}$ distribution with the error bars depicting the 1$\sigma$ mean square variation.  The distribution within the error bars is very similar to the observed distribution of red points. This also indicates that a lot of stars could hide in the upper left corner at high pericenter distances and large eccentricities. We have repeated this test, adopting a Rayleigh distribution which is characterized by the probability

\begin{equation}
P(e_{\star})=\left( \frac{e_{\star}}{\sigma} \right) \times  \exp \left[ 0.5 \times (1-\left( \frac{e_{\star}}{\sigma} \right)^2) \right]
\end{equation}

\noindent The best fit in this case corresponds to $\sigma =0.6$ (yellow points). Despite this additional free parameter it cannot fit the data
as well as the random distribution. Finally, the open points show an isothermal distribution P(e$_{\star}$)$\sim$e$_{\star}$ which cannot fit the data at all.

\section{Summary}
We demonstrate that the Galactic center S-star cluster shows a narrow, exponential dependence of pericenter distance r$_p$ on (1-e$_{\star}$), with e$_{\star}$ as the orbital eccentricity. The G-clouds follow the same correlation, indicating that they might be of similar origin or have experienced a similar secular evolution. If scattering mechanisms were important, the fact that the G-clouds were not disrupted would indicate that they might have an invisible, compact central source \citep{scoville2013, peissker2021}. We then derive the likelihood to determine orbital parameters for S-stars observationally and show that small values of (1-e$_{\star}$), combined with large pericenter distances r$_p$ require a larger observational time span than what is available at the moment. This explains why no stars have yet been found in this region. It might be populated with yet undetected S-stars. On the other hand, the orbital parameters of stars characterized by $\log$ (r$_p$/AU) $<$ 1.57+2.6(1-e$_{\star}$) should be easy to determine. That no stars exist in this zone of avoidance indicates that whatever dynamical processes shaped the S-star cluster kinematics did not favor this region. Considering the observational biases and the zone of avoidance, we find that the S-stars follow a random sampling of $\log$ r$_p$ with (1-e$_{\star}$). Our results could provide valuable constraints for theoretical models of the origin of the S-star cluster and its kinematics. For example, we will show in an accompanying paper that this correlation arises naturally from the perturbation of an intermediate-mass companion (Zheng et al. in preparation) without invoking any tidal-disruption-of-binary hypothesis \citep{hills1988, chu2023}.

\begin{acknowledgements}
This work was supported by the Excellence Cluster ORIGINS which is funded by the Deutsche Forschungsgemeinschaft (DFG, German Research Foundation) under Germany's Excellence Strategy - EXC-2094 - 390783311.
\end{acknowledgements}

\bibliography{paper.bib}{}
\bibliographystyle{aasjournal}

\end{document}